\documentclass{mem}
\usepackage{natbib}
\usepackage{txfonts}
\usepackage{balance}
\usepackage{graphicx}
\usepackage[a4paper]{hyperref}
\idline{76}{4}
\begin{document}
\def\teff{$T\rm_{eff }$}
\def\kms{$\mathrm {km s}^{-1}$}

\title{
Toward a Census of Variable Stars in Northern Local Group Dwarf
Irregular Galaxies
}

   \subtitle{}

\author{
J. Snigula\inst{1} 
\and C. G\"ossl\inst{2}
\and U. Hopp\inst{1,2}
}

\offprints{J. Snigula}

\institute{
Max Planck Institut f\"ur Extraterrestrische Physik, 
Giessenbachstr., D 85748, 
Garching, Germany
\and
Unisternwarte M\"unchen
Scheinerstr. 1,
81679 M\"unchen, Germany,\\
\email{snigula@usm.uni-muenchen.de}
}

\authorrunning{Snigula }

\titlerunning{Variable Stars in LG Dwarfs}

\abstract{ Dwarf galaxies in the local group provide a unique
  astrophysical laboratory. In particular, they allow us to probe
  pulsating (and other) variable stars in low-metallicity environments
  with abundances below that of the SMC. Our observing program,
  described in detail by C. G\"ossl's contribution, yields a large
  number of intrinsically bright variable stars that can serve as
  probes of the stellar population and star formation history of these
  galaxies. Most prominent are pulsation variables like Miras (LPVs)
  and $\delta$ Cep stars, but we also find other types of variable
  stars, e.g. RV Tauri stars, irregular red variables etc.  We present
  a preliminary census for the three galaxies DDO 216, Leo A and GR8.
   }
\maketitle{}

\section{Sample and Data Reduction}
\begin{figure}[]
\resizebox{\hsize}{!}{\includegraphics[clip=true,angle=-90]{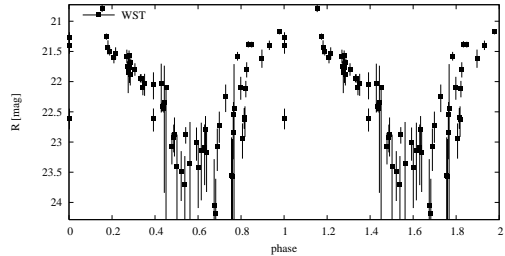}}
\caption{\footnotesize
  Light-curve for a LPV with 255 days period in the pegasus dwarf
  galaxy (DDO 216).
}
\label{lpv2}
\end{figure}

We selected a sample of six local group irregular dwarf galaxies. So
far the observations were carried out in the $R$- and $B$-Band
sparsely sampling a three year period starting with test observations
in 1999.  This part of the data set consists of approximately 80
individual epochs per galaxy and is sensitive to long period variable
stars with periods up to $\sim 500$ days.  Additional observation in
the $R$, $B$ and $I$-Bands were obtained during three observing
campaigns at the 1.23 m telescope on Calar Alto densely sampling three
two week long periods. These observations provide a ground for a
search for variable stars with shorter periods ranging from $\sim 1.5$
days up to $\sim 10$ days. The depth of each epoch is roughly $22.5$ mag
in the R-Band.
\begin{figure}[]
\resizebox{\hsize}{!}{\includegraphics[clip=true,angle=-90]{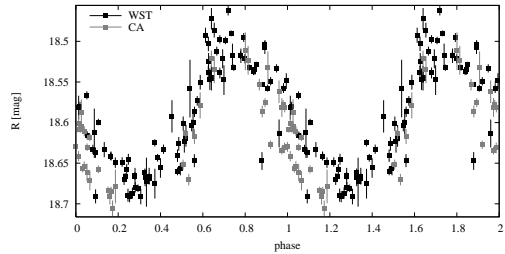}}
\caption{\footnotesize
  Light-curve of an LPV in Leo A with a period of 74.9 days. The black
  points are observations from Mt. Wendelstein, the grey points from
  Calar Alto.}
\label{lpv1}
\end{figure}
\begin{figure*}[!t]
\resizebox{\hsize}{!}{\includegraphics[clip=true,angle=-90]{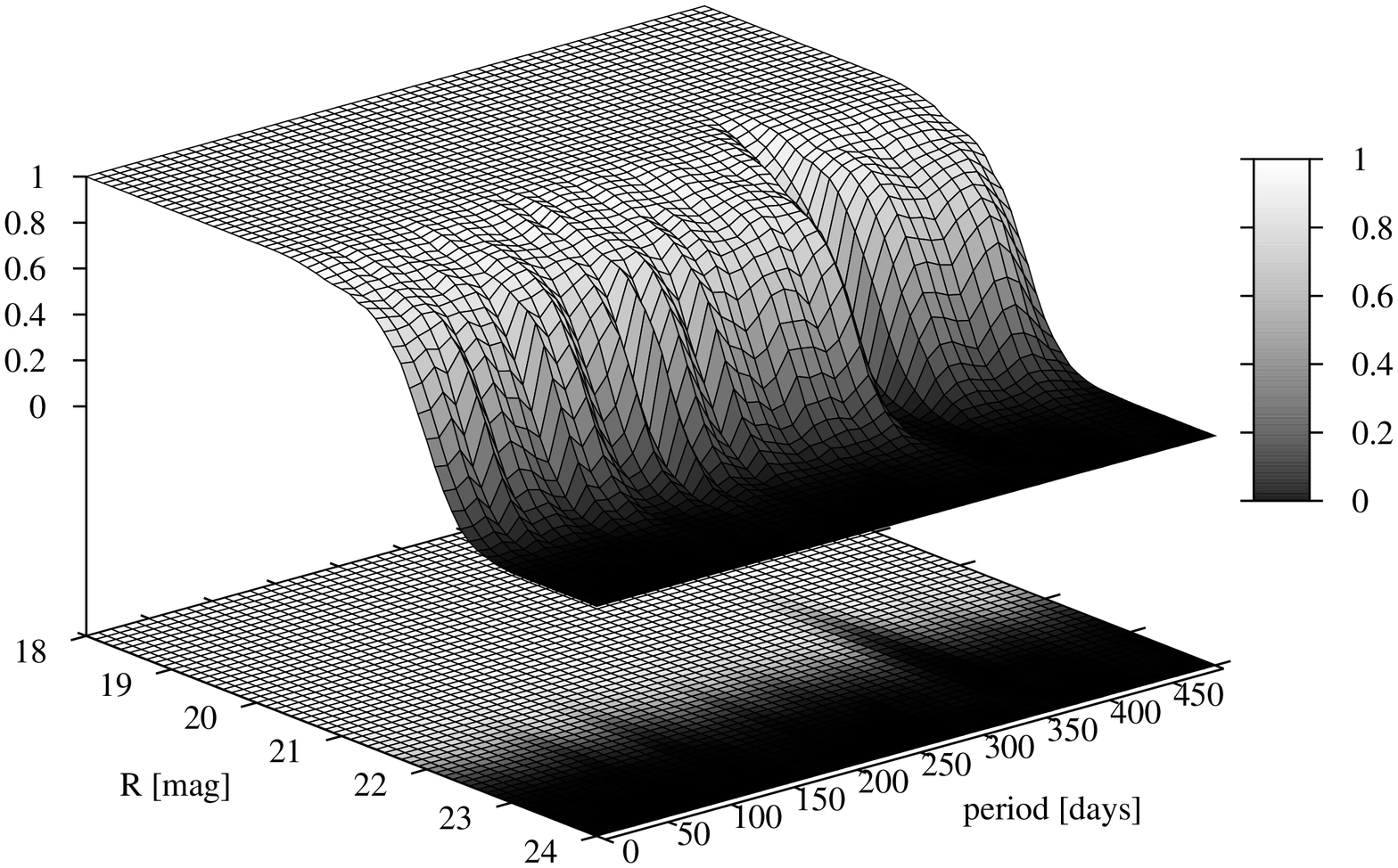}
  \hspace{0.1\textwidth}
  \includegraphics[clip=true,angle=-90]{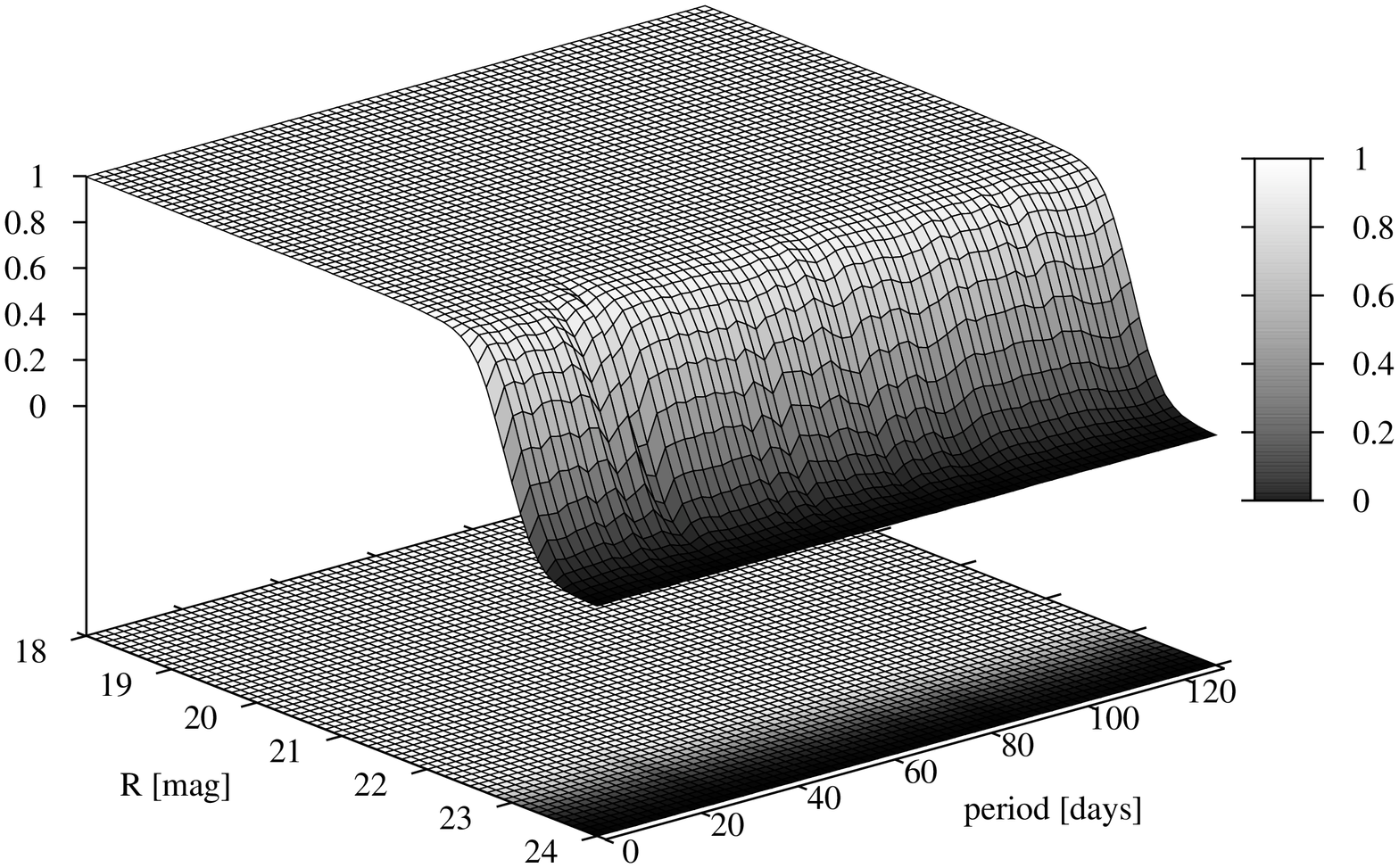}
}
\caption{\footnotesize
  Completeness simulation for variables with a cosine shaped
  light-curve and an amplitude of 1 mag, using the Lafler-Kinman
  statistic (left panel) and the Lomb algorithm (right panel). 
  The break-in for periods of about one year in the left figure, can
  be explained by our half-year observing window.}
\label{comp_lk}
\end{figure*}
The acquired data were bias-subtracted, flat-fielded and
cosmic-rejected, at the same time propagating the error of each pixel.
Consequently the images from one night were astrometrically aligned to
a common reference frame and combined with individual weights
proportional to the S/N.  For each epoch, a difference image against a
common deep reference frame was created using the Alard algorithm
\citep{Alard} implemented by \citet{Claus}, still propagating the
individual pixel errors. In a final step these difference images were
convolved with a stellar PSF. The short period variables were detected
using an implementation of the Lomb algorithm \citep{Scargle}. For the
LPVs the Lafler-Kinman \citep{Lafkin} statistic was applied.
Lightcurve examples for LPVs are shown in Figs.~\ref{lpv1}
and~\ref{lpv2}.

\begin{table}[]
  \caption{\footnotesize
    Preliminary census of the detected variables.}
  \label{census}
  \begin{center}
    \begin{tabular}{l|cc}
      & $\delta$ Cep & LPVs \\
      \hline
      Leo A   & $14$ & $16$\\
      DDO 216 & $10$ & $47$\\
      GR 8    &  $1$ &  $3$
    \end{tabular}
  \end{center}
\end{table}

\section{Completeness Simulations}
To obtain a measure of the completeness of the resulting catalogue of
variable sources, we carried out extensive simulations covering the
complete set of relevant parameters: magnitude, period and amplitude.
The tests were conducted using a sample of nearly 900 artificial
sources. As light-curve shapes both a cosine and a sawtooth were used,
testing both an ideal, as well as a worst case. With exception of the
limitation for one year periods the simulations show no notable
deficiency (Fig.\ref{comp_lk}).

\section{Preliminary Variable Census}
So far, we finished analyzing three galaxies from our sample. Table~\ref{census}
gives a short overview of the amount of detected variables.

The next step will be the final reduction and evaluation of the
remaining dwarf galaxies, and a scientific analysis of possible SFHs
of these, using the found pulsation variables and the completeness simulations.

\bibliographystyle{aa}

\end{document}